# Exploiting Data Reduction Principles in Cloud-Based Data Management for Cryo-Image Data

Kashish Ara Shakil[1], Ari Ora[2], Mansaf Alam[1]* and Shabih Shakeel[3]*

Affiliation: [1]Department of Computer Sciences, Jamia Millia Islamia, New Delhi, India

[2]Aalto University, Espoo, Finland

[3]Institute of Biotechnology, 00014 University of Helsinki, Helsinki, Finland

*Corresponding authors: malam2@jmi.ac.in, shabih.shakeel@helsinki.fi

## Abstract

Cloud computing is a cost-effective way for start-up life sciences laboratories to store and manage their data. However, in many instances the data stored over the cloud could be redundant which makes cloud-based data management inefficient and costly because one has to pay for every byte of data stored over the cloud. Here, we tested efficient management of data generated by an electron cryo-microscopy (cryoEM) lab on a cloud-based environment. The test data was obtained from cryoEM repository EMPIAR. All the images were subjected to an in-house parallelized version of principal component analysis. An efficient cloud-based MapReduce modality was used for parallelization. We showed that large data in order of terabytes could be efficiently reduced to its minimal essential self in a cost-effective scalable manner. Furthermore, on-spot instance on Amazon EC2 was shown to reduce costs by a margin of about 27 percent. This approach could be scaled to data of any large volume and type.

## Introduction

Cloud computing is the use of computing resources as a service over the internet. Cloud computing involves varied amount of data coming from different customers across geographically dispersed locations (1). This data can be in form of unstructured data such as videos and images or semi structured data from emails, structured data, flat files, or data coming from databases such as Sybase and legacy data (2). Due to the advantages and features of cloud computing such as on demand self-service, pay per use, rapid elasticity and dynamic scalability, there has been a paradigm shift to cloud by organizations. One of the most noteworthy advantage of using cloud computing is its ability to easily scale out to clusters of any size(3). Since data in cloud is held by a third party there are several issues such as security and management of data coming from variety of sources such as social networking sites, bank processes, images, audios and videos that needs to be taken into account. Therefore, the organizations are now looking for concrete techniques for managing data in a cloud based system.

Exponential growth in biological data, has led to difficulties in management and storage (4). Consequently, it has led to a huge gap between data generation and computing capabilities. In the field of life science, a newly established lab usually have a limited start-up grant and in such cases utilizing the resources offered by cloud computing could be a cost-effective, and productive way of starting a lab. However, many times such labs produce highly redundant data which would add up to the storage cost over the cloud and in case such data needs to be processed, then it would add to the processing over the cloud as well. Our experience in working with images obtained from electron cryo-microscopy (5) projects show that the amount of data produced could easily run into terabytes especially with the advent of direct electron detectors and automation. For a new established lab, storage and processing of such data could be prohibitive. In such cases, lab-over-cloud could be highly useful. In addition, such data many times contains a mix of good and bad images; and removal of bad images immediately before long-term storage over the cloud could be cheaper and computationally effective.

There are many strategies available to reduce data redundancy. One of the most successful methods is based on principal component analysis (PCA) where majority of 'similar' data gets segregated from the one with high difference. It is a multivariate analysis technique that allows analysis of variables that are correlated (6). PCA on a data matrix allows extraction of patterns which are dominant in the matrix through loading plots and complementary scores (7). It is mathematically dependent on singular value decomposition (SVD) and eigen decomposition of positive semi definite matrices. The modeling performed using PCA can be divided into four parts: data, score, loading and residual (8).Unfortunately, implementing PCA for a large data-set of a terabyte or even few giga-byte size data could be inefficient. Therefore, storage and processing of such big volumes of data is the requirement of numerous users.

Hadoop MapReduce is a de facto standard for processing large datasets in both the industry and academics (9). Therefore, we have utilized map-reduce modality available as part of Hadoop (10) framework for PCA. Performing PCA using map-reduce greatly enhanced its performance. In this work we implemented PCA in Amazon EC2 cloud environment. This paper demonstrates an approach how cryo image data can be efficiently reduced, stored and processed in a cost effective manner on a public cloud. Results demonstrate the scalability and robustness of the proposed workflow. Furthermore, application of dimensionality reduction techniques further reduces the storage costs involved. Following are the contributions of this paper:

- Proposal of a workflow pipeline for storage and processing on cryo-image data on cloud

- Assessing the effectiveness of solution on a public cloud in terms of cost, scalability and robustness

**Scalable computing on Public cloud: Amazon EC2:**

Amazon EC2 (11) is the section of Amazon Web services responsible for offering cloud services in form of rented virtual machines known as Amazon Machine Image or an instance. It has been utilized for storage of biological data such as GenBank (12), Ensembl (13) and 1000 Genomes (14). It offers its users with a quick scaling up and scaling down of the compute resources depending upon their requirements. This change in scalability offered by Amazon EC2 is highly elastic in nature. Many organizations such as Novartis (15), Bristol-Myers-Squibb (16) and researchers have now begun to use Amazon EC2 in order to meet their computing needs due to an extensive set of benefits offered by it such as high reliability, auto scaling, security and robust networking.

Amazon offers different instance types depending upon the customer's need (17). The instances differ on the basis of CPU, storage, memory and networking providing its customers with wide number of options to choose from. These instance include T2 instance which provide burstable performance, M4 (latest generation general purpose instance), M3, C4 (compute optimized), C3, R3 (memory optimized), G2 for graphics applications, I2 (storage optimized and high I/O instances), and D2 (dense storage instances). These instances come in different sizes thus providing its users with an elastic scalability depending upon the changes in their workload behavior.

The usage of Amazon EC2 leads to many economical benefits. The users have to pay a very nominal charge for the compute resources they use. At present there are three different purchasing models available. The first one is an On-demand instance model where the customers pay for the period of time they use the resource on an hourly basis with no upfront costs. The second model is the "reserved instance model", and is suitable for applications with a predictable workload usage. It is cheaper than the On-demand model but requires initial upfront costs. The third model is the Spot Instances model that allows customers to use computing resources in an hourly manner with no upfront costs. This model can at times be cheaper than the On-Demand model. In this model, the prices vary depending upon the load on the data centers. In this model, users have to bid for the maximum price they are willing to pay for a particular instance and if the current spot price is less than the bid, then the instance is started otherwise terminated.

**CRYO Image DATA**

The resolution revolution in electron cryo-microscopy field has led to many labs to adopt this technique to solve the structures of proteins in order to understand their function. Image data collection using this microscopy is fully automated and the microscope is run 24x7. Additionally, the images are collected as movies with ~7-50 frames per movie. Automation and movie mode collection has caused a tremendous surge in the amount of data collected in each microscopy session and can easily run into terabytes per session. However, not all data collected is used in image processing, some of the data contains bad ice regions, junk particles and carbon surface which needs to be discarded.

For a newly established laboratory in electron cryo-microscopy, investing in data storage could be a huge cost that's why cloud storage seems lucrative. Unfortunately, every byte of data stored in the cloud adds to the cost, therefore, storing only the essential data which goes into image processing is needed. Here, we propose to run PCA on the electron cryo-microscopy data in the cloud-based environment to identify the "to be discarded" data so as to save the data storage cost over the cloud.

Use of cloud computing in managing cryo-image data can help the new users to immediately access a high performance computing cluster and can also help existing life sciences labs to increase their productivity. For example, in developing countries, there are limited resources available to researchers. Thus, lack of funds and resources can be limiting criteria for the users. These problems can be easily solved by leveraging the advantage of public cloud facilities offered by Amazon. They provide an immediate and cost-effective access to unlimited computing resources irrespective of their geographical locations.

**MapReduce**

Parallel computations can run on top of 'parallel frameworks', to make computations efficiently and in a fault tolerant manner. Table I gives a comparison between the different parallel approaches:

**Table I**

**Comparison between different parallel approaches**

| Approach | Advantage | Disadvantage |
| --- | --- | --- |
| Message Passing Interface (MPI) (18) | Allows programmers to create parallel programs | Complicated software development |
| Batch processing systems (like Condor) (19) | Can run simple parallel programs | Can't run complicated algorithms |
| MapReduce (20) | Automatically handles job scheduling, distributed aggregation and fault tolerance. | Not efficient for running computations on small datasets. |

MapReduce is the center of Hadoop paradigm. Hadoop has been used in several bioinformatics applications, leading to cloud-based bioinformatics service (21). It has two main components Hadoop distributed file system (HDFS) and MapReduce. The advantage of using MapReduce as a programming model is that it offers massive scalability (20). The functioning of MapReduce can be divided into two steps: Map phase where a set of tasks are broken into individual elements and then the processing is carried out followed by reduce phase which aggregates the output generated from individual map functions and represents them as a consolidated result.

## Material and Methods

For the purpose of our experimentation, the On-Spot pricing option was adopted and instance type r3.8x large was used. The r3 instance is a memory intensive instance. Each of the instance types provides up to 32 virtual CPUs (vCPU). The vCPUs provided by Amazon are slightly different from the actual physical CPUs i.e.2 vCPUs =1physical CPU. Thus, in the instance r3.8xlarge used in this experimentation 16 actual CPUs were used per node. Each node was able to process up to 10GB of cryo image data.

In order to find out variations in data PCA was performed. At first image dataset which was of the total size of 100 GB, was taken from the Electron Microscopy Pilot Image Archive Repository (EMPIAR) with EMPIAR-10012 and consisted of 100 images with each image of pixel size 7420 by 7676. Hence, each image had a total of $144(n^2)$ dimensions. After this for each image in the image dataset conversion of image data to vector space was done. The vector space data was then loaded into a distributed computing system by creating a data store. Covariance was then calculated for the data set using the covariance MapReduce function. Covariance matrix is represented by 'C'

$$C = A*A^T \qquad (1)$$

$$A = \{\Phi_1, \Phi_2 ..., \Phi_M\} \qquad (2)$$

$\Phi$ are columns of matrix and dimension of A= $N^2*M$ where M is the number of images (number of columns) and N is the number of rows

C will be of $N^2*N^2$ dimension

The vector space was then projected onto the eigenspace. In order to meet this end, we first calculated the correlation matrix from the covariance matrix by finding standard deviations and then normalized it to correlation form.

Correlation = C/ (s*s') where s is the square root of the diagonal of the covariance matrix i.e. s = sqrt (diag (Covariance)) and finally, PCA was performed by us using svd on the Correlation matrix. Algorithm1 ALGOCryoDimensionReduce was used by us for computing PCA of the image data.

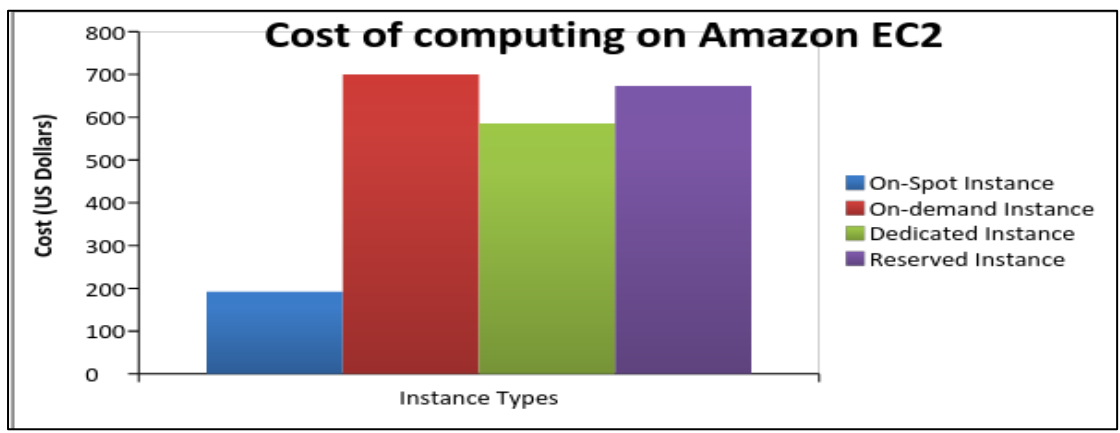

**Fig 1:** Cost of running 2Tb of cryo image data using different pricing schemes

**ALGORITHM 1**
**ALGOCryoDimensionReduce**
**Input**: Indata : Input Cryo Imagedata
**Output**: Pcacoef : Reduced Matrix after performing PCA
1. **Begin**:
2. Outdata = CryoMapReduce (Indata, Cryocovariancemapper, Cryocovariancereducer) /* Running CryoMapreduce on input data to apply map and reduce functions*/
3. C = CryoCovariance( Outdata ) /*covariance is calculated on the data output from Cryomapreduce */
4. Sq = sqrt(diagonal( C)) /*square root of the diagonal elements of the covariance matrix obtained from previous step is calculated */
5. Correlation = C/Sq*Sq' /* Computing Correlation matrix through standard deviation and then it is normalized to correlation form */
6. Pcacoef ← svd( Correlation ) /* singular value decomposition is used for performing PCA on the correlation matrix */
7. return Pcacoef /* return the coefficient matrix obtained from step 6*/
8. **End**

## Results & Discussion

Cryo image data of different sizes were taken from the cryo image data repository, EMPIAR (EMPIAR IDS:EMPIAR-10012. beta-galactosidase, Bartesaghi A 2014 PNAS.) and transferred to Amazon for data storage. The cost of storing and processing 2 terabytes of data was calculated using different pricing schemes available on Amazon as shown in Fig 1. The prices were then compared and it was then observed that cheapest processing costs were obtained with 'on spot pricing' scheme compared to 'on demand pricing', 'dedicated' and 'reserved instances costs' scheme. Use of on-spot pricing scheme led to up to 27 percent reduction in costs Thus, 'on spot pricing' was used for data storage and processing throughout this work.

**Cost-Effectiveness**

We tested the tradeoff between size of data versus the computational cost. We found out that the cost of processing increases with increase in size of data. This increase in cost approaches a linear value as shown in Fig 2.

Instances were bid for $ 0.96 per hour using US West (Oregon) Region. We successfully conducted PCA on all the images and were able to pull-out good images out of the bad ones. Fig 3 shows a scatter plot for the data after PCA was performed.

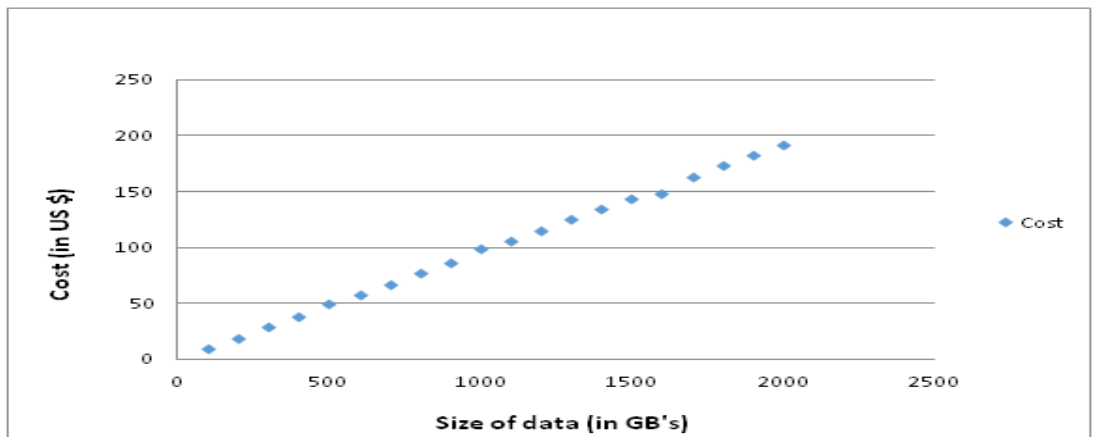

**Fig 2:** Tradeoff between size of data and Cost using On-Spot pricing

Furthermore, by conducting PCA, if for example, we are able to reduce the size of data from 2 TB to 1.5 TB, then we will end up saving approximately 50$.

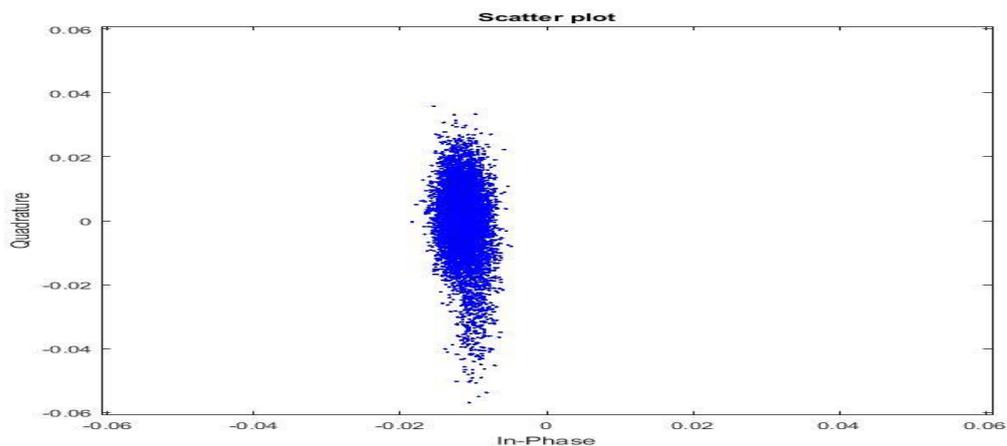

**Fig 3:** Result obtained after performing PCA on Cryo image data

With the above procedure, we propose a new data management pipeline as shown in Fig 4, where the data exported to cloud gets sorted via a fast-performing PCA, so as to retain only the useful data which help in reducing storage cost. The steps of the proposed method are as follows:

*Step 1: Ingest cryo-image data into cloud storage such as AWS master node or S3 bucket.*
*Step 2: Apply ALGOCryoDimensionReduce*
*Step 3: Pick the relevant or good images*
*Step 4: Store data back on cloud storage*

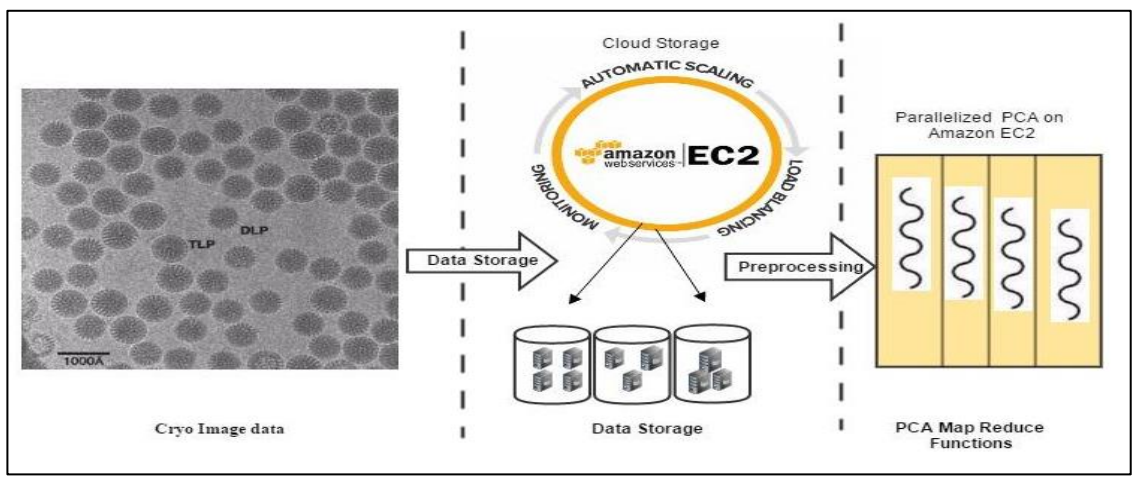

**Fig 4:** Proposed Workflow Pipeline

The experiments were conducted using PCA algorithm implemented through Hadoop MapReduce on Amazon EC2. 200 instances of r3.xlarge instance types were utilized in our experiments. The advantage of using MapReduce framework is that this approach is scalable and therefore can handle data of huge volume and variety in a dynamic manner. At first, a data store for Cryo-microcopy images is created. The data store for this experiment contains about 2TB of unstructured data or image data. After the data store has been created, a MapReduce PCA algorithm is conducted which performs PCA in a parallel fashion.

**Scalability**

We tested the scalability for our procedure as it is one of the salient features of MapReduce. 305 MB of data got processed via PCA in 40 seconds. Similarly, we were able to extend this approach for data of 2 TB and were able to perform this computation on 2TB of data in 60.8 hours. However, the boot time for the instances has not been considered in the calculation of processing time.

**Robustness**

Utilization of cloud computing modality ensures robustness as the system offers high availability by replicating the data across several nodes. In the case of failure, if a node is down another node can be started off immediately.

Thus, a public cloud such as Amazon EC2 can be an alternative to that of high-performance clusters used in traditional computing laboratories. Distributed computing clusters can be easily created, managed and hosted on a public cloud platform like Amazon EC2 using utilities like StarCluster (22). It provides a quick mechanism to set up a cluster of machines to perform processing on data on top of distributed frameworks. This work will help in the spread of cloud computing to be used as a tool for cryo-image analysis apart from analysis of other kinds of data in bioinformatics. It also saves on the computation cost and time involved in the startup of a research laboratory for data involving heavy computations.

**Conclusions**

This work proposes a novel data management pipeline for exploiting data reduction principles in a cloud-based environment. Use of cloud computing is a cost-effective storage and management solution for life science laboratories. Cryo-image data generated from electron microscopes have been used in our study. The data was reduced through principal component analysis; on a public cloud i.e. Amazon EC2. PCA was parallelized using MapReduce programming paradigm. The results show that minimum cost can be incurred through On-spot pricing mode on Amazon EC2 and saves about 27 percent costs. Further, by performing PCA on about 2 TB of data we can additionally reduce this cost by 50$.The advantage of storage and management using this approach offers a scalability and ability to handle data of any volume and variety.

This data reduction modality can be utilized in several other fields where large data is involved besides life sciences such as geospatial data and sensor data management. Furthermore, the proposed workflow pipeline can be easily modeled into an 'off-the-shelf' computing environment for processing such large datasets.

In future we plan to extend this work to GPU enabled cloud techniques, as it provides significant parallelism and will lead to further upto70 percent reduction in costs.

**Acknowledgment**

This work was supported by a grant from "Young Faculty Research Fellowship" under Visvesvaraya PhD Scheme for Electronics and IT, Department of Electronics & Information Technology (DeitY), Ministry of Communications & IT, Government of India.